\documentclass[12pt]{amsart}
\usepackage{amsmath}
\usepackage{amsfonts}
\usepackage{amssymb}
\usepackage{graphicx}%
\providecommand{\U}[1]{\protect\rule{.1in}{.1in}}
\newtheorem{theorem}{Theorem}

\newtheorem{remark}[theorem]{Remark}
\newtheorem{rmk}{Remark}

\newtheorem{prf}{Proof}

\newcommand{\NOT}[1]{}
\newcommand{\pa}{\par\medskip}

\title{Let the Mathematics of Quantum Speak: Allowed and Unallowed Logic}
\author{Eliahu Levy}
\address{Department of Mathematics, Technion -- Israel Institute of
Technology, Haifa 32000, Israel} \email{eliahu@math.technion.ac.il}

\keywords{quantum physics, logic aspects in quantum, logic allowed and unallowed, closely following the mathematics/formalism, commutative and noncommutative algebras, truth-values, logic of possibilities vs.\ yes-no, system of possibilties vs.\ the `actual world', Boolean algebras, probability, Hilbert space, operators, closed subspaces, orthogonal projections, density matrices, magnitudes, observables, almost-commutative, quasi-classical, operations defined pointwise, measurements, Copenhagen quantum `recipe',\
decoherence, determinism}

\date{}

\begin{document}

\begin{abstract}
Some notes about quantum physics, an interpretation if one wishes, are put forward, insisting on `closely following the mathematics/formalism, the `nuts and bolts of what quantum physics says'. These, basically well-known, issues seem to highlight some rather bold points about the `logic' aspect in quantum physics, necessarily restricting when and which logic may be admissible. And one may understand why that path is hardly followed in the literature.

The mathematics/formalism of quantum, compared with classical, physics, may be fairly basically characterized by non-commutative algebras replacing commutative. These classically appearing, in fact, in dealing with systems of possibilities (say, all possible planetary motions under gravity of which one is the actual one).

In particular, contrary to too common usage, the quantum non-commutativity should make it impossible to simply `transcend' the `system of possibilities' aspect into a `yes-no' logic essential for an `actual world'. One may have the latter only in a `haven' of approximately commutative algebras of `quasi-classical macroscopic observables', and moreover that `yes-no actual world' would plainly be an `extra ingredient' to the base quantum theory itself.
\end{abstract}

\renewcommand{\thesection}{}

\maketitle

\tableofcontents

\subsection{Introduction}\label{s:int}
Quantum physics, in a similar way to Probability and (mostly historically) Differential and Integral Calculus, has been riddled with major arguments about how to interpret it, with multifaceted paradoxes to tackle. That despite the fact that all three cases stand on a clear-cut established mathematical formalism -- a term applied to contrast with `deep' essence -- which tells precisely when formulas and modes of solution to problems are valid and what the (mathematical) theory predicts experimentally.

But let us follow closely the mathematics/formalism, the nuts and bolts of what quantum physics says. Then, basically well-known considerations seem to highlight some rather bold points about the `logic' aspect in quantum physics, necessarily restricting when and which logic may be admissible (see below). And one may understand why that path is hardly followed in the literature.

That mathematics/formalism can be fairly viewed as characterized, compared with classical physics, by non-commutative algebras replacing commutative. These classically appearing, in fact, in dealing with systems of possibilities (say, all possible planetary motions under gravity of which one is the actual one).

In particular, contrary to too common usage, the quantum non-commutativity should make it impossible to simply `transcend' the `system of possibilities' aspect into a `yes-no' logic essential for an `actual world'. One may have the latter only in a `haven' of approximately commutative algebras of `quasi-classical macroscopic observables', and moreover that `yes-no actual world' would plainly be an `extra ingredient' to the base quantum theory itself.

I recapitulate here some points in \cite{Levy-Occ}.

\subsection{The Advent of Quantum Physics}
Adhering thus to the mathematics/formalism, one may fairly say that \textit{the advent of quantum physics came when observations and experiments forced physicists to replace the classical commutative algebra related to a system of possibilities (to wit, the algebra of bounded magnitudes) by a non-commutative algebra (to wit, the algebra of the (complexified) bounded observables)}.

Which turned to be just the right thing so that things fit.

Bluntly, as per the formalism, our case is basically closed -- we know how to oblige to Nature the Autocrat.  Mathematically=formalistically speaking, \textit{to understand the base of quantum physics we have to cope with that non-commutativity}.

That classical commutative algebra refers, of course, to a system of `possible worlds'. Very often in classical physics `the actual world' is approached as one `state' or `possible world' in a system of possibilities. Thus there are the laws of planetary motion under gravity, which allow many possible scenarios, one of which is the actual case.

And mathematically we have the algebra of scalar \textbf{magnitudes}, say complex-valued (usually with some qualification such as `bounded') -- these having, in general, different numerical values at different possible worlds. Define the algebraic operations there \textit{pointwise}, i.e., say, addition of magnitudes by adding the values they take at each possible world. They form an \textit{algebra} -- a set where operations like addition, multiplication, multiplication by a complex scalar and (complex) conjugation are defined and satisfy standard requirements.

In such system the logic is a `logic of possibilities' -- logical statements (`events' in the parlance of probability theory) cannot be said to be `true' or `false'. Rather, they are true in some possible worlds and false in others. They form a \textbf{Boolean Algebra}, bigger than just the truth values $\mathbf{2}=\{\mathbf{'True'},\mathbf{'False'}\}$.

The algebra of magnitudes is obviously \textbf{commutative}, here meaning that the commutative law for multiplication $ab=ba$ holds (while mathematics has important examples for non-commutative algebras -- notably the algebras of matrices or of linear operators in linear spaces).

And as very common in such mathematics, one may start from either of the structures: the set of `possible worlds'; the boolean algebra of `events'; the algebra of magnitudes, and retrieve = define the others. In this sense each can be the primary notion.

In particular, the possible worlds (here with the role of `states', even `points'), can be retrieved as functionals on the algebra, i.e.\ scalar-valued functions, assumed to be \textit{homomorphisms}, i.e.\ to respect all the operations%
\footnote{Considering, of course the, say complex, scalars as naturally also an algebra with operations.}
 -- let us refer to these as \textbf{hom-functionals}.

Indeed, identify a `point' $a$ with the \textit{hom-functional whose value at a magnitude $\phi$ is the value $\phi$ takes at $a$}!

And with assumptions that often hold, one proves sorts of theorems%
\footnote{I have in mind theorems like Wedderburn's theorem that a finite-dimensional \textit{semisimple}
non-commutative algebra is a cartesian product of simple algebras, in the commutative case the latter must
be the scalars; Gelfand's theorem that a commutative $C^{\ast}$-algebra is always (isomorphic to) the algebra
of continuous scalar functions on a Hausdorff compact topological space; that a commutative von Neumann
algebra is necessarily $L^\infty(X)$ for $X$ some measurable space; etc.}
which say that \textit{giving a (properly qualified) set of points is `the same' as giving a (qualified) commutative algebra: the points are (in one-one correspondence) with the hom-functionals on the algebra as above; and the algebra is (isomorphic) to that of (qualified) magnitudes on the points, with operations defined pointwise}.

Bluntly speaking, when that is the case \textit{a commutative algebra is just another way to say a (totally classical) many-world scenario}.

And keep in mind that all that refers to a \textit{system of possibilities} where a statement (= event) cannot be said to be simply true or false! That is, of course, untenable as a final conclusion. Science, which has to tell us about `the real world' should assert: yes or no.

Still, when commutative we know that the algebra can very often be viewed as embodying a set of possibilities/possible worlds/states, if you wish those being (identified with) the hom-functionals on the algebra, of which, in the commutative case, there are very often plenty, enough to determine and build the algebra.

And in each particular one of these `possible worlds' statements \textit{are} `true' or `false'!

So in the classical/commutative scenario we may try to `fool ourselves' by pretending
to have in mind some \textit{particular} possible world, with statements having truth values, `yet letting it vary'.

Returning to the quantum case, we then have a non-commutative algebra, which, in analogy to the commutative case, should refer to a set of possibilities, where statements/events certainly are not just `true' or false.

The usual way that the quantum non-commutative algebra is described is, of course, as the (in fact, von Neumann) algebra  $B(\mathcal{H})$ of bounded operators on a Hilbert space $\mathcal{H}$. But for us the algebra has the first place, the vectors in the Hilbert space etc.\ might be said to serve a role of indices (cf.\ the discussion of `states' below).

One may say that the non-commutative algebra makes, primarily, a description of \textit{a new and strange logic} on \textit{the system of possibilities}.

In that, \textit{an accompanying Boolean-like algebra is missing}. Indeed, one cannot base the logic, as in the commutative classical case, on operations among the logical statements/events themselves.%
\footnote{From the point of view of the algebra, these should be the analogs to the numbers $0,1$ ($=\{\mathbf{'True'},\mathbf{'False'}\}$) among the scalars, the numbers $a$ satisfying $a^2=a$. For magnitudes that equation would characterize a magnitude which takes values in $\{0,1\}$,
the characteristic function of a set $E$ of points = an event, (i.e.\ the magnitude taking value $1$ in $E$ and $0$ in its complement).

For the non-commutative scenario, to wit the algebra $B(\mathcal{H})$ of operators in a Hilbert space,
these are \textit{orthogonal projections}, members $p$ of the algebra satisfying $p^2=p$ (adding the requirement $p^{\ast}=p$ -- Hermitian -- the analog automatically holding for $0,1$-valued), orthogonal projections on closed subspaces in the Hilbert space, thus in one-one correspondence with these.}
Boolean operations among events = subspaces = projections such as conjunction and disjunction are naturally partially defined (just for compatible events, equivalently commuting projections).%
\footnote{One may be tempted to extend the definition of union and intersection to non-compatible events as the sum and intersection of the relevant subspaces of the Hilbert space. Note, however, that these depend highly non-continuousely on the subspaces -- when an `angle between subspaces' turns to zero the intersection and sum spaces both `jump', which seems to definitely disqualify that.}

The logic should rather be thought of as \textit{given by the non-commutative algebra itself}, whose operations are always defined. (Thus there is no harm if the original algebra, which serves to define the logic, contains only \textit{bounded} observables although in general observables are unbounded.)

\begin{remark}
So, noting the mathematical difference in properties, we see when this `novelty' may be dealt in some analogy with the classical/commutative case, and when things are totally different.

Referring, in particular, to \textit{probability measures}, the transit to quantum is rather smooth. In the commutative case probability measures on the `points' = `possibilities' would, from the point of view of the algebra of magnitudes, be given by \textit{positive linear functionals on the (commutative) algebra of magnitudes mapping the unit element of the algebra to $1$} -- the functional which gives to a magnitude its expectation and in particular to the characteristic function of an event $E$ the probability of $E$.

By analogy, the role of probability measures will be played, in the non-commutative case too, by positive functionals with value $1$ at the unit element $\mathbf{1}$.%
\footnote{To be mathematically correct, in the infinite-dimensional case, only positive linear functionals belonging to the \textit{predual} of the von Neumann algebra (such as $B(\mathcal{H})$).}
Thus when the algebra is $B(\mathcal{H})$ -- the (von Neumann) algebra of the bounded operators in a Hilbert space, these will be the \textit{density matrices}, i.e.\ positive semi-definite operators $\tau$ with trace $1$. Relative to such $\tau$, an element of the algebra, i.e.\ an operator $A\in B(\mathcal{H})$, will have the expectation
$\text{tr }(\tau\cdot A)=\text{tr }(A\cdot\tau)$. In particular an event -- projection $p$ (i.e.\ $p^2=p$ and $p^{\ast}=p$) will have the probability $\text{tr }(\tau\cdot p)=\text{tr }(p\cdot\tau)$ which will be in the interval $[0,1]$.
\end{remark}

\subsection{States? What Can or Cannot Be}
Recall again, that crucially, the above non-commutative algebra=logic should define just the quantum counterpart of the \textit{logic of possibilities}.

In the classical/commutative case one often has the algebra as just another way to say `a set of (possible) worlds', retrieved as the set of hom-functionals from the algebra to the complex scalars.

Then certainly only such states/possible worlds could (as a genuine world with usual logic obviously must) endow statements/events of the system of possibilities with a truth-value `true' or `false', so that these events occur or not.

In the system of possibilities itself these are just elements of a Boolean algebra, and saying that they occur or not is meaningless. The logic consists of manipulating them, if one wishes as members of that Boolean algebra,

The same should hold for the non-commutative quantum system of possibilities. To go beyond possibilities -- to be able to assert some statement as `true' or `false', we would need to refer to a state/possible world.

A full, genuine such state/possible world should thus be a \textit{bona fide} hom-functional. Then every observable will get its value and every event will get the value $1$ or $0$ $=$ `true' or `false'. And in the commutative case we had them plenty and sufficient.

Yet alas, one knows mathematically that \textit{for a non-commutative algebra that is out of the question}.
Such hom-functionals to the scalars are scarce and insufficient, sometimes nonexistent, in a non-commutative algebra -- it definitely does not make a many-world scenario (totally classical!) as the commutative case did.

The (pure) states spoken of in quantum physics, given by wave-functions = elements of a Hilbert space up to multiplication by a scalar (thus by definition always abundant) are something else, definitely not where statements are `true' or `false'. In the above sense they are not states at all.

These are events characterized by being minimal (one may use the word `atomic' in the mathematical sense), i.e.\ they have no proper sub-events. Equivalently, there is only one `probability measure' supported in each of them. Physically, they give the maximum specificity that one can have, what in the classical/commutative case had characterized single states/possible worlds.

But a general event need not either happen in them (contain the atomic event) or not happen (its complement contains the atomic event),  equivalently their probability distribution gives to a general event values in the interval $[0,1]$ different from $0$ or $1$. In particular, future events have just probabilities with respect to such mathematically atomic `states'.

In this sense these pure states, which give the maximum specificity that one may conceive, are like general events -- sets of states -- and general probability distributions on them in the (deterministic) classical systems. Thus one should not wonder that this maximum specificity does not determine the future (yielding only probabilities).

Again, as we have emphasized, the quantum non-commutative structure should refer just to a \textit{system of possibilities}, thus it is meaningless to say that statements are true or false, events occur or not. What the Theory does is just manipulating them, as members of the non-commutative logic -- to wit subspaces  of the Hilbert space (equivalently orthogonal projections) using its clear-cut formalism.

Otherwise put, as we have seen, the mathematics/formalism definitely tells us that:

The Quantum Theory does of course richly speak about particles, fields, physical systems etc.\ and about statements = (events = subspaces of the Hilbert space), such as `the electron is here or there'; `it has this or that property'; `the system is (or was) in that state'. \textbf{But to stress again}: these being actors in the non-commutative logic of \textit{the system of possibilities}, the Theory aptly manipulates them according to its clear-cut formalism, but \textbf{they in no way happen, are true (or not happen, are false)}, in spite of the tempting wording -- asking that is meaningless.

That was the case also with the commutative/classical scenario. Yet as said above we then could, on the back of our mind, think that we are dealing with a set of `possible worlds', so they are true or false \textit{for each one of them} and judge ourselves as `having in mind some particular possible world, yet `letting it vary'. As we saw, for the quantum/non-commutative even that small consolation is denied us.

\subsection{Our Actual World (an Extra Ingredient)}
But, in this quantum picture, we still have to recover our actual world -- where events occur or not -- and our usual logic. Otherwise put, we have to locate the as if \textit{haven}, inside Quantum, where statements \textit{can} be true or false.

It seems clear where to find that. We need a commutative sub-algebra, and something like that presents itself: \textit{the algebra of the macroscopic, quasi-classical observables which almost commute}. (In fact, even its members should be viewed as only approximate -- they cannot be handled in greater precision than the uncertainty that makes them commute.)%
\footnote{One may object that our approximate notions here are hardly given a mathematical definition. Maybe one can try some mathematically rigorous ways to do that. I still do prefer this kind of discourse, which seems to be within the scope of the ways of physics, to picking some mathematically correct definitions, but otherwise arbitrary from the point of view of the situation at hand itself.

Roughly, that approximate discourse should mean that `one can say something if and only if one stays within the allowed approximation'.}
These, and the events they define, are what we have in our old classically behaving world.%
\footnote{Note that the evolution of a system in Time is defined by conjugation with imaginary exponents of the energy (Schr\"odinger's equation), So the energy, itself quasi-classical, cannot exactly commute with other quasi-classical observables, otherwise there would be no time evolution there. Similarly with the momentum which induces variation in space.}

And our actual world is described by an \textit{approximate} hom-functional on this (approximately commutative) quasi-classical algebra. It will give values $0$ or $1$ to projections/events/closed subspaces of the Hilbert space \textit{which belong to this quasi-classical algebra}, (again, a $p$ satisfying $p^2=p$ must map by a hom-functional to a scalar doing the same!) i.e.\ truth values `true' or `false' to these quasi-classical statements. Only these events occur or not in our actual world, and only with them we can use our usual logic.

Indeed, as far as the approximation goes, having an (approximately) commutative algebra, we naturally have a `many-world' scenario -- the worlds correspond to (`are') all possible (approximate) hom-functionals from this algebra to the complex scalars.

Which our world is one of, its choice being \textbf{an extra ingredient} -- from the point of view of the quantum theory other ones could have equally been chosen. (One might wonder whether we could not deduce everything in our actual world from `probability close to 1' arguments. This seems not to be the case. It seems that in many cases quantum fluctuations have been magnified to macroscopic consequences, making many different outcomes each with small probability, of which just one is asserted in the actual world. And moreover there are so many details in our actual world that seem entirely erratic.)

\subsection{Paradoxes?}
And a great origin of paradoxes in thinking about quantum theory is our reluctance to obey its `strange' logic.

As we must contend, we have, on the one hand, the quantum theory, for us the non-commutative algebra, a description of a new and strange logic on the \textit{system of possibilities}, which the Theory aptly manipulates as members of the non-commutative logic -- mathematically, operators, orthogonal projections in the Hilbert space -- using its clear-cut formalism. But saying there that a statement = (event = subspace of the Hilbert space) is true or false would be meaningless. That will have meaning only per our actual world, and just for an event belonging to the quasi-classical almost commutative algebra. Only then we have our logic.

But the Quantum Theory does speak about particles, fields, physical systems etc., and one is so tempted to say, in its frame, that `the electron \textbf{is} here or there'; `it \textbf{has} this or that property'; `the system \textbf{is} (or was) in that state' as if events there happened or not, which the non-commutative logic of the `system of possibilities' forbids.

And then one runs straight into paradoxes.

\subsection{Some Interplays Quantum -- Classical, as per Our Actual World}
The approximate nature of our actual world is usually unnoticed by us, since we ourselves come from this approximate world.

But it will limit the number of different events (statements) that we can meaningfully conjunct or disjunct -- without totally leaving the quasi-classical almost commutative algebra. Thus it limits the number of things -- amount of information -- that we can speak about (to something like a `mundane' action measured in Planck's Constant -- something like $10^{34}$);

It limits the amount of time to the past or future that can have meaning for our actual world -- because the non-commutativity with observables transformed by Hamiltonian evolution, although small and negligible for our mundane intervals of time, becomes big for enormous intervals, hence one cannot include presumably quasi-classical quantities pertaining to enormously distant times or distances -- something like multiples of mundane times or distances by the ratio of a mundane action to Planck's constant -- in the same approximate commutative algebra;

All that making our physical actual world \textit{finite to a delimited extent}.

One may even say that as per the `actual world', the infinite space or time models used in physics serve, in this respect, a similar role as the infinite plane of coordinates in which a map includes the grounds of a city.

Of course, we can investigate non-quasi-classical systems only by making them bear on our almost-commutative quasi-classical world, i.e.\ by measuring them.

Moreover, our world is protected from `stray non-commutativity', such as carrying conclusions of former measurements to the future via the Hamiltonian (Schr\"odinger's equation) evolution, by \textit{decoherence} \cite{JoosZeh} \cite{Zurek} which will wipe out any such conclusions, preserve only what is quasi-classical and (almost) commuting and thus create the separating wall between the quasi-classical and the truly quantum worlds.

Any Schr\"odinger cat (or `Schr\"odinger physicist or mathematician, for that matter) is either in the quasi-classical domain, hence one may assume in principle that in our actual world the question: is (s)he alive? is settled, or is in the truly quantum domain, where superpositions are routine, but can then be investigated by us only via measurements.

Usually, the quasi-classical world is governed by the deterministic laws of classical physics, to be derived, in principle, from the quantum theory. But the fact that everything is approximate has consequences. Thus when the deterministic classical equations are chaotic we have true non-determinism in the quasi-classical system: between assertions about far enough time-moments one may have only probabilistic relations. Another case of non-determinism comes from measurements and measurements-like phenomena, where `truly quantum' elements bear on the quasi-classical world.\pa

\subsection{The `Measurement Problem'}\label{s:mea}
One may say that the original version of quantum theory has put itself in a physics laboratory.

Atomic systems are prepared (by measurements, selecting the cases where a favored outcome occurred, say electrons moving in a specific direction) and measured. Such systems (more exactly, their way of preparation) always have a state, described by a wave function, or even by a density matrix (in case the preparation had a classically random element). Doubtlessly, these states and their measurements are governed by Copenhagen quantum theory, as repeated in numberless textbooks, with collapse of the wave function and everything.

All that works perfectly well as long as one sticks to the laboratory scenario, in which the division of labor, so to speak, among the quantum system, the experimenter(s) and the laboratory apparatus goes unquestioned. (A great part of the problematic nature of the Schr\"oedinger Cat example in Copenhagen quantum theory seems to stem from the unnatural role of a sentient being as the quantum system.)

So let us look at the `measurement problem' in view of what we noted above.

Consider a proverbial quantum measurement. A quantum system is prepared, by making a measurement and taking only the cases with suitable outcomes (say, `electrons that move in a certain way'), Then, maybe after some development in the quantum system, another measurement is made, then, maybe, more development and measurements.

We assume that each measurement distinguished between all vectors in a basis of the Hilbert space of the quantum system to be measured (there is no classical randomness).

We, of course, live in the quasi-classical world (in fact in our actual world -- an extra ingredient), where in our actual world the measuring apparatuses recorded results of the measurements. Here, contrary to a quantum `non-commutative system of possibilities', assertions are true or false -- it is true at the measurements gave these results.

Anyhow, an essential ingredient of that being a measurement, is recording that the quantum system that one measures now was indeed prepared (by a former measurement) as required. The total Hilbert space must have room for keeping records of all these.

In fact, as per the quasi-classical almost commutative algebra, roughly speaking, each subsequent measurement tensors (qua Hilbert space, qua algebra of operators) the already recorded results with the results of the new measurement.

And as per our actual world -- all these quasi-classical events either occur of not.

Note also that, as per the quasi-classical almost commutative algebra, if one assumes knowing that a quasi-classical event = subspace occurred, one might work only in (`condition to', in the probability parlance)  that subspace as the Hilbert space.

\subsection{The Copenhagen Recipe Follows Smoothly}
Then, maybe to one's surprise, it seems that (restricting ourselves to `allowed logic' as above) the Copenhagen recipe follows smoothly -- no extra assumptions.

This is because mathematically, each subsequent measurement, by adding to the quasi-classical record, basically makes our entire Hilbert space tensor with the Hilbert space of the new (quasi-classical) record. In writing the probability distribution -- density matrix -- that we should take, always in a basis that (approximately) diagonalizes the quasi-classical observables, each entry $w_{ij}$ of the density matrix will be replaced, by the above tensoring, with a sub-matrix $W_{ij}$ with trace $w_{ij}$, (a positive Hermitian sub-matrix $W_{ii}$ in the case of a diagonal element $w_{ii}$), and, \textit{speaking in our world -- only there could we assert that anything is true or false} we condition to the quasi-classical event of the specific result of the new measurement (true in the actual world), hence pick the corresponding particular entry in the diagonal of that sub-matrix.

For the next measurement this entry itself is to be expanded into a sub-matrix, etc. Note the remarkable fact that every subsequent measurement as if imposes its pure state as the new density matrix after the conditioning -- we can `forget' what density matrix we used before. In other words, we must compute as if each measurement had induced a collapse of the state of the measured quantum system (which has nothing to do with the group of time-shifts -- Hamiltonian evolution -- which of course, always acts by Schr\"odinger's equation).

\appendix

\section{Some Parallels Relativity -- Quantum}
One can find similarities between the introduction of Quantum Theory and that of Special and General Relativity:

As the latter denies special status to particular frames of reference, so the former denies special status to particular commutative subalgebras, equivalently special status to particular bases of the Hilbert space; and as the latter makes the mathematics technically more demanding but conceptually simpler, uniting space with time, mass with momentum and energy and space-time with gravity, so the former unites the observables with the action symplectic structure into the non-commutative algebra.

In this sense, quantum theory is here to stay like any other basic part of physics. A future theory may perhaps encompass it in a more fundamental/comprehensive theory, but not obviate it, as Newtonian physics was not obviated, just made a special case $=$ approximation of our more fundamental theories in certain (important) settings.

And as, in the context of Relativity Theory, some people try to `sacrifice' Lorentz invariance and revert to special space-time frames of reference, so in the quantum contexts one may try to have a preferred basis. It seems to me that the thrust of a `hidden variables' attempt consists in a special status that, for some reason, would have to be given to a special basis, equivalently to a special commutative subalgebra, making the general non-commutative algebra a subordinate structure, so to speak (which even basic considerations of, say, angular momentum make rather unlikely).

\section{Epilog: Confessions?/Apologies?}

This contribution, relying primarily on mathematics, claims that as per the non-commutative quantum Theory (= Logic), which so aptly manipulates the events -- orthogonal projections in the Hilbert space, such as: `the electron is here or there'; `it has this or that property'; `the system is (or was) in that state', according to its clear-cut formalism, still definitely does not allow stating them as statements -- true (or false).

That applying, in particular, to saying that a system (or the universe) \textbf{is} in that or that state (= has that or that wave-function).

But, as said in the beginning of \S\ref{s:int}, in quantum (as in the other examples: probability and -- mostly historically -- differential and integral calculus), physicists and mathematicians know very well when formulas and modes of solution to problems are valid and what the (mathematical) theory predicts experimentally. In the quantum case they usually work in the frame of the Copenhagen `recipe' -- see the beginning of \S\ref{s:mea}.

Where they very well say that, say, that and that \textbf{is} the state (=wave function) of a system --
`deducing' that from results of measurements on it, in particular those used to `prepare' it.

And I deliberately try to distance as possible from things with `philosophical' flavor, like: are they philosophically, ontologically, correct in saying so, thus `violating' my standpoint? The same as in mathematics one may choose to introduce or not $\infty$ as a `number', as long as mathematical rigor is kept, ignoring such ontology/philosophy. Of course, my standpoint will show that `their' way may be extended only so far, lest they run into inconsistencies and paradoxes.

It seems that the need for `interpretations' arises primarily from clear uneasinesses and paradoxes. And there that `philosophical' flavor reigns supreme, such as what `exists', in contrast to: just something one may talk about.

For example, the whole point of Everettian many-worlds seems to disappear if one does not care whether existing or imaginary, conceivable.

I talked about the Copenhagen \textit{'recipe'} -- also to distinguish from and avoid the philosophy flavor of Bohr's as much as possible.

I emphasized that one \textit{listens to the mathematics} of quantum. Not creating a new one.
I very much dislike when one creates special physics and mathematics theories \textit{just for the sake of the philosophical satisfaction}, as is so much encountered in `interpretations'.

Note also that I have talked only about logic -- statements, `events' (in the parlance of probability theory), and not about objects and properties, predicates, about which the logic speaks -- electrons. systems, the Universe etc.

(Do such `exist'? Two quotes from Wittgenstein: The world is what is the case (i.e.\ the statements that are true) -- the first line in the \textit{Tractatus}; The table I see is not made of electrons.)

In a system of possibilities (all possible planetary motions under gravity of which one is the actual one; the theory of groups) such objects must be meaningful across the possibilities, in fact \textit{by decree}: the planet Jupiter; the unit element of the group.

Ordinarily a universe of discourse comes with its proper logical language. In our world `water' or `life' or `tuberculosis' are meaningful as they exist. What these terms would mean in a different world
with different properties is dubious -- or decided by decree.

\end{document}